\documentclass[preprint,12pt]{elsarticle}

\usepackage{amssymb}
\usepackage{subfigure}
\usepackage{color}

\journal{Physica A}

\begin{document}

\begin{frontmatter}



\title{Distinguishing manipulated stocks via trading network analysis}


\author[label1]{Xiao-Qian Sun\corref{cor1}}
\ead{sunxiaoqian@software.ict.ac.cn}
\author[label1]{Xue-Qi Cheng\corref{cor1}}
\ead{cxq@ict.ac.cn}
\author[label1]{Hua-Wei Shen\corref{cor1}}
\ead{shenhuawei@software.ict.ac.cn}
\author[label1]{Zhao-Yang Wang}
\cortext[cor1]{Corresponding address:\\
      No.6, South Road, Kexueyuan, zhongguancun, Institute of Computing Technology,Chinese Academy of Sciences, Beijing 100190, China.\\
}
\address[label1]{Institute of Computing Technology,
        Chinese Academy of Sciences, \\
        Beijing 100190, China}

\begin{abstract}
Manipulation is an important issue for both developed and emerging stock markets. For the study of manipulation, it is critical to analyze investor behavior in the stock market. In this paper, an analysis of the full transaction records of over a hundred stocks in a one-year period is conducted. For each stock, a trading network is constructed to characterize the relations among its investors. In trading networks, nodes represent investors and a directed link connects a stock seller to a buyer with the total trade size as the weight of the link, and the node strength is the sum of all edge weights of a node. For all these trading networks, we find that the node degree and node strength both have tails following a power-law distribution. Compared with non-manipulated stocks, manipulated stocks have a high lower bound of the power-law tail, a high average degree of the trading network and a low correlation between the price return and the seller-buyer ratio. These findings may help us to detect manipulated stocks.

\end{abstract}


\end{frontmatter}

% \linenumbers


\section{Introduction}

Dynamic behavior of stock markets has attracted much academic and industrial attention. Studies on dynamic behavior are facilitated by the availability of numerous historical financial data accumulated with the development of stock markets. The financial data can be roughly classified into two categories: time series of financial variables and detailed transaction data which contain the information of trader identities and their transactions.

For the first category of financial data, the traditional research approach is to employ the probability distribution functions in order to analyze statistical properties of various variables~\cite{Stanley07,Stanley03,Stanley2001,Stanley00,Stanley99,Lee07,Zhang07,Zhou09,Gu09,Qiu09,Zhou08,Zhou10,Wang01,Sun10} and to employ correlation functions to study the cross correlation among different variables~\cite{Podobnik08,Stanley09,Zhou2008,Ren10,Podobnik10,Stanley2003,Lee04}. A power-law distribution has been found in many variables, such as price fluctuations, trading volume and the number of trades. In~\cite{Stanley03}, a model is proposed to explain these empirical power-law distributions. Network analysis provides another method for studying the financial market~\cite{Onnela04}. For example, researchers analyze the networks of listed companies~\cite{Qiu10,Piccadi10,Mantegna99} or stocks index fluctuations~\cite{Li06,Li07,Cai10}. Community structure and hierarchical structure are two properties of complex networks~\cite{Fortunato10,Cheng09,Shen09,Shen10}. The community structure is shown in two real-world financial networks, namely the board network and the ownership network of the firms of the Italian Stock Exchange~\cite{Piccadi10}. Mantegna find a hierarchical arrangement of stocks by investigating the daily time series of the logarithm of stock price~\cite{Mantegna99}.

Unlike the data of the first category which characterize the macro-level properties of a stock market, the second category of financial data describe the specific transactions among investors from a micro-level perspective. From transaction data, a trading network can be constructed, in which traders are mapped into nodes and each transaction relation corresponds to a directed link. Kyriakopoulos et al. investigated the statistical properties of the transaction network of all major financial players within Austria over one year~\cite{Kyriakopoulos09}. They found that the transaction network is disassortative and that eigenvalue analysis is helpful for detecting abnormal behavior of financial players. Tseng et al. conducted a market experiment with 2,095 effective participants in order to explore the structure of transaction networks and to study the dynamics of wealth accumulation~\cite{Tseng10}. They found that the transaction networks are scale-free and disassortative. In addition, they find that the wealth distribution follows Pareto's law. Wang et al. constructed a trading network using the real trading records from Shanghai Future Exchange and found that future trading networks exhibit such features as scale-free behavior, a small-world effect, hierarchical organization, and a power-law betweenness distribution~\cite{Wang11}. Furthermore, in ~\cite{Zhouweixing10}, Jiang and Zhou first studied the trading networks of all traders of one stock in the Chinese stock market. They found that the trading networks comprise a giant component and have power-law degree distributions and disassortative architectures.

The endeavors on the network of investors indicate the micro-level mechanism provides a promising way to shed light on the relationship between the macro-level statistical properties and the micro-level trading behavior. In this paper, to study the manipulation of stock, we focus on the micro-level of trading behavior. Specifically, we investigate the investors' trading behavior using the transaction data of up to 108 liquid stocks. We first analyze the statistical properties of the trading network and find that the networks constructed from every stock have power-law degree distributions. We further find that the node strength distributions of the trading networks also have power-law tails. Finally, through comparison between manipulated stocks and non-manipulated stocks, we find that manipulated stocks have a high lower bound of the power-law tail, high average degree of the trading network and low correlation between the price return and seller-buyer ratio. These findings provide a promising way of detecting manipulated stocks.

The paper is organized as follows. Section 2 describes the data set. In Section 3, statistical properties of the trading network are analyzed. In Section 4, the features of manipulated stocks are analyzed and then compared with those of non-manipulated stocks. Finally, the conclusions are given in Section 5.

\section{The data sets}

In a stock market, a trader submits bid/ask orders to the electronic trading system when he/she wants to sell/buy shares. In the trading system, the ask orders are sorted in ascending order of price and the bid orders are sorted in descending order. Through matching a bid order and an ask order according to a certain rule of price/time priority, the trading system generates a transaction record. All of the transaction records constitute the transaction data which are the raw data used in this paper. These matched orders are often called executed orders.

In this paper, we study the transaction data of 108 stocks traded on Shanghai Stock Exchange and Shenzhen Stock Exchange during the whole year 2004. The transaction data for the 108 stocks contain 44,333,930 transaction entries, which involve 11,686,740 unique trader accounts. Each entry consists of the date and time, the unique number for the transaction, the buyer ID, the seller ID, the volume and price. Among all these 108 stocks, eight stocks had been manipulated by some investors and these manipulated stocks are revealed by China Securities Regulatory Commission(CSRC) for trade-based manipulation. In addition, for five of the eight manipulated stocks, the manipulated period persists through the whole year of 2004 and the manipulated period for the remaining three stocks is from Jan 2004 to Sep 2004. The basic description of the data sets is reported in Table \ref{table:1}.

\begin{table}[htb]
\caption{Basic statistics of the data sets. \emph{N} is the total number of stocks. $<N_{t}>$ is the average number of traders, and $<N_{e}>$ denotes the average number of entries.}
\label{table:1}
\newcommand{\m}{\hphantom{$-$}}
\newcommand{\cc}[1]{\multicolumn{1}{c}{#1}}
\renewcommand{\tabcolsep}{1pc} % enlarge column spacing
\renewcommand{\arraystretch}{1.2} % enlarge line spacing
\small
\begin{tabular}{@{}lllllll}
\hline
\cc{$Data set$} & \cc{$N$} & \cc{$<N_{t}>$} & \cc{$<N_{e}>$} &\cc{$Period$}\\
\hline
Non-manipulated & \m100 & \m111,215 & \m416,719 & \m Jan 2-Dec 31\\
Manipulated & \m5 & \m29,564 & \m119,838 & \m Jan 2-Dec 31 \\
Partially manipulated & \m3 &\m139,130 & \m687,597 & \m Jan 2-Sep 3 \\
\hline
\end{tabular}\\[2pt]
\end{table}

\section{Stock trading network}

Generally speaking, a stock market comprises of two main entities, namely list companies and investors. Stock trading behavior occurs between buyers and sellers, and all the trading behavior forms a directed network, which is called the stock trading network. The stock trading network provides a straightforward representation for characterizing the trading relationship among institutional or individual investors. It is believed that the study of the stock trading network may provide some insights into the relationship between the price fluctuation and the collective trading behavior.

\subsection{Construction}

\begin{figure}[!htb]
  \centering
  \includegraphics[width=25pc]{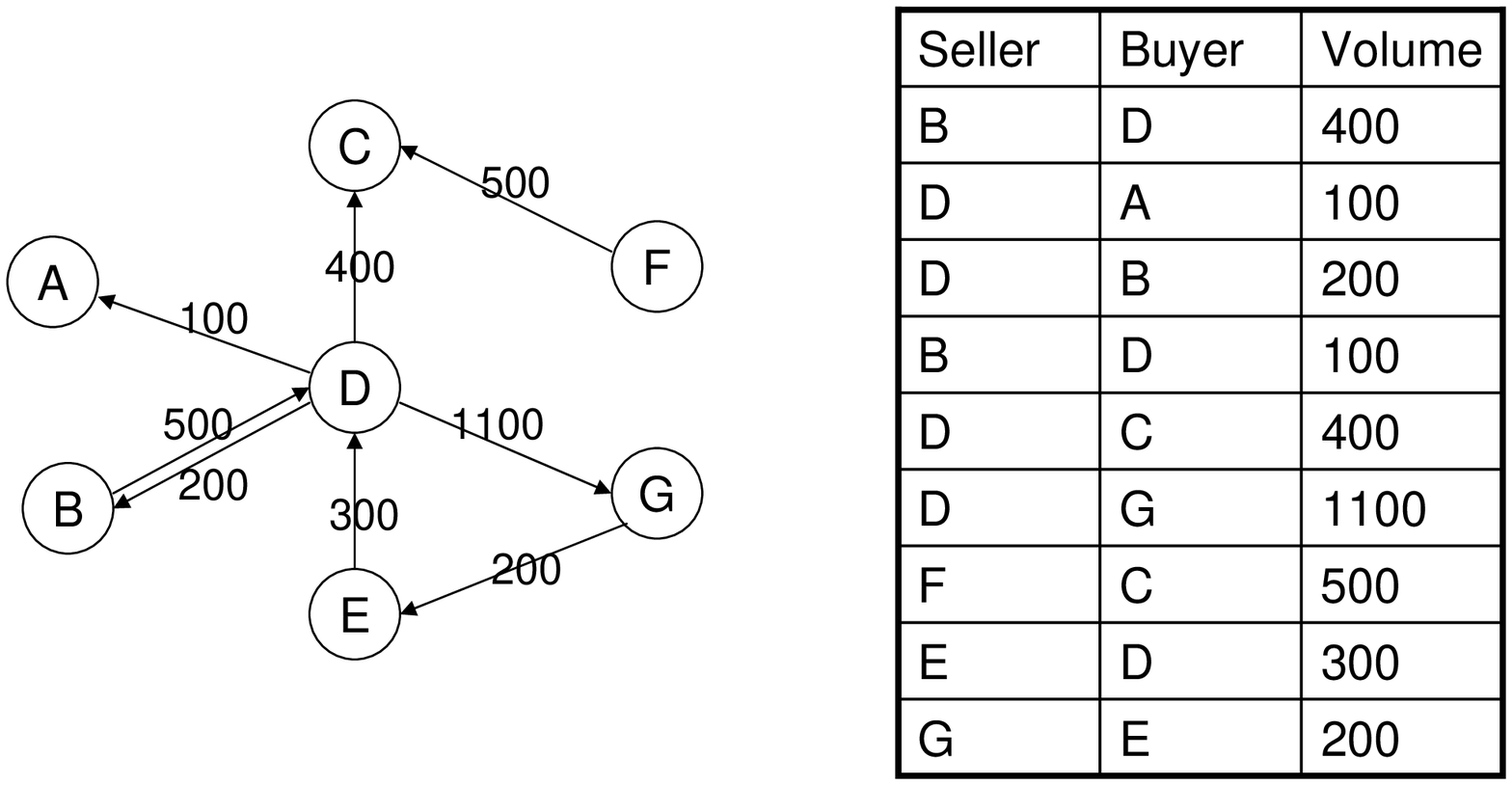}\\
  \caption{A schematic example of a stock trading network(left) constructed from nine transaction records(right).}\label{example}
\end{figure}

\begin{figure}[!htb]
  \centering
  \includegraphics[width=25pc]{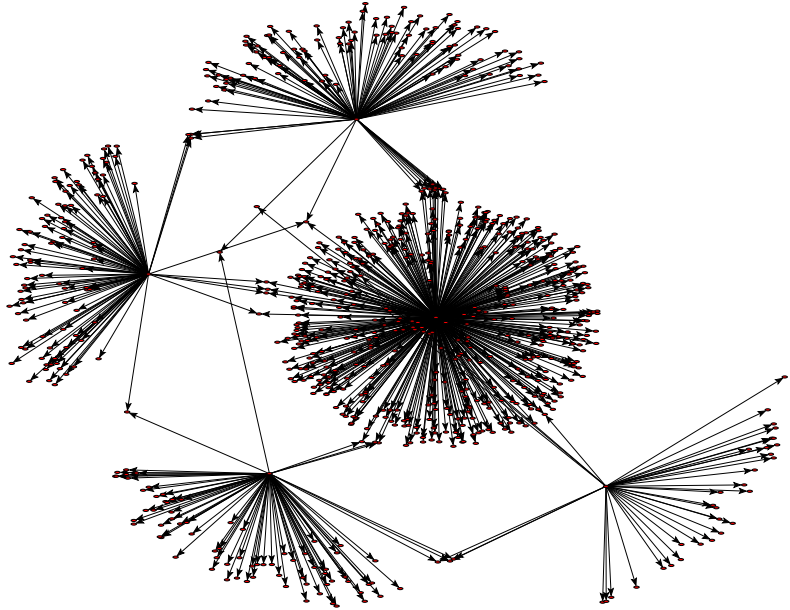}\\
  \caption{A real stock trading network for a certain stock on Jan 8 2004, and there are 726 nodes and 748 links.}\label{transaction}
\end{figure}

We first describe the construction process of the stock trading network according to the transaction data. For each specific stock, we construct a trading network using the transaction records in a whole year. Specifically, the nodes of the trading networks correspond to buyers and sellers involved in the transaction data of the stock considered. For each transaction record, a directed edge is constructed pointing to the buyer from the seller. The volume of the transaction record is taken as the weight of an edge. In this way, there may exist multiple edges between two nodes if more than one transaction occurred between the same seller and buyer. Note that the directed and weighted edges reflect the flow of shares from the seller nodes to buyer nodes and the flow of cash in the opposite direction. Without losing such a physical meaning, we merge the multiple edges with the same seller and buyer nodes into one weighted edge. The final trading stock network is a directed and weighted network. Figure \ref{example} illustrates the construction of the stock trading network using a toy example. To give an intuitive understanding, Figure \ref{transaction} gives a stock trading network of a real non-manipulated stock in our data set.

In \cite{Zhouweixing10}, a stock trading network is constructed using the transaction data for each trading day. However, as pointed out in \cite{Mookerjee99}, the stock market suffers weekend, holiday and monthly effects. To alleviate such effects, we aggregate the transaction data of the whole year into a sole trading network instead of constructing one trading network for each trading day. In addition, no isolated node exists in trading networks since each transaction involves two traders, a seller and a buyer.


\subsection{Degree distribution}
The degree of a node in a network is the number of connections that the node has to other nodes. Since the edges of the trading network are directed, the network is characterized by two different degrees, the out-degree, which is the number of outgoing edges, and the in-degree, which is the number of incoming edges. The degree distribution function $P(k)$ of a network describes the heterogeneous properties of nodes. The distribution of outgoing edges $P_{out}$(k) signifies the probability that an investor sells shares to \emph{k} investors, and the distribution of incoming edges $P_{in}(k)$ is the probability that an investor buys shares from \emph{k} investors. We investigate two probability distributions for each stock during the whole year period. For the investor \emph{i}, the more nodes it connects to, the more active the corresponding investor is.

\begin{figure}
\centering
\subfigure[]{
    \label{fig:mini:subfig:a}
    \begin{minipage}[b]{0.5\textwidth}
    \centering
    \includegraphics[width=\textwidth]{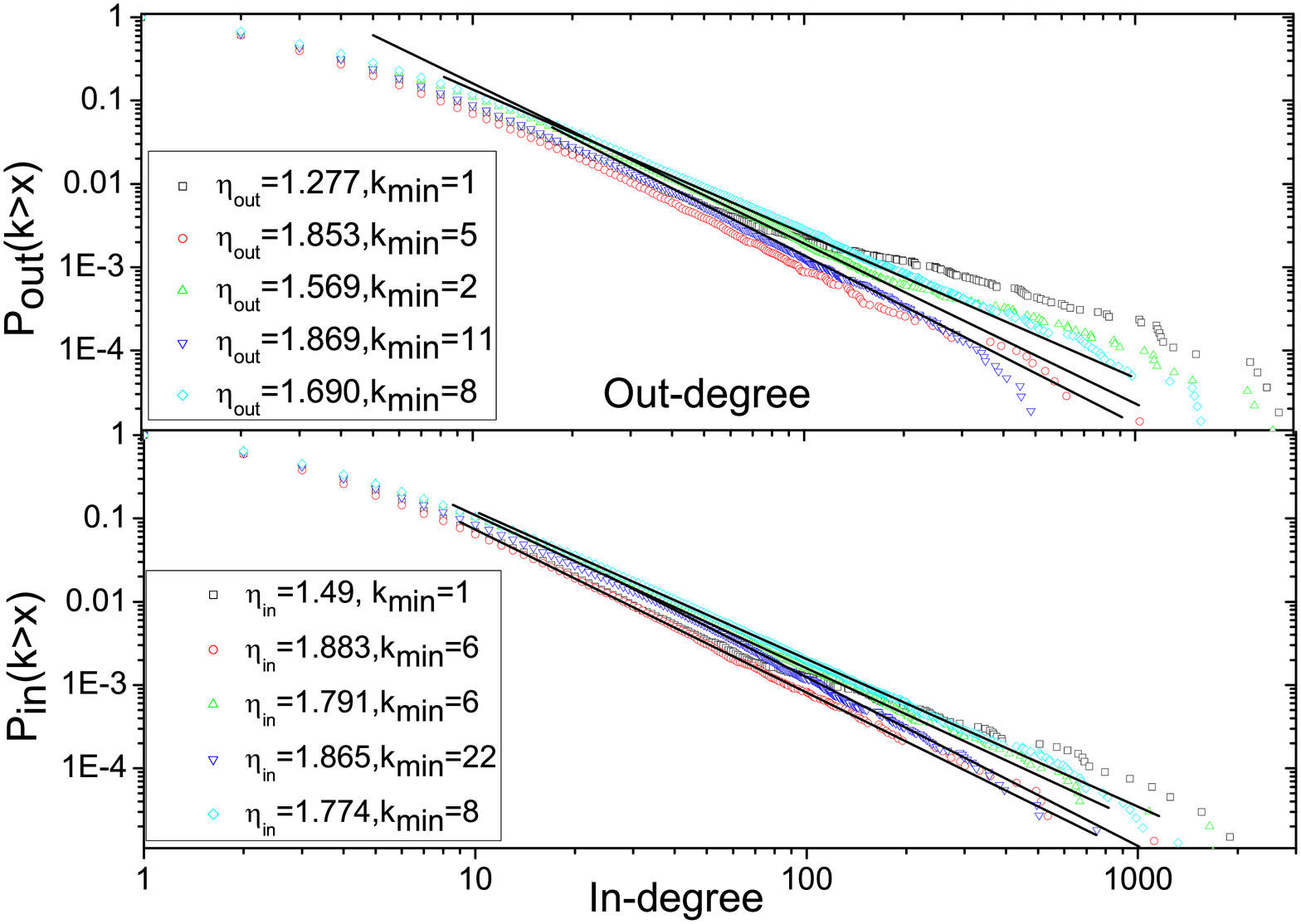}
    \end{minipage}}%
\subfigure[]{
    \label{fig:mini:subfig:b}
    \begin{minipage}[b]{0.5\textwidth}
    \centering
    \includegraphics[width=\textwidth]{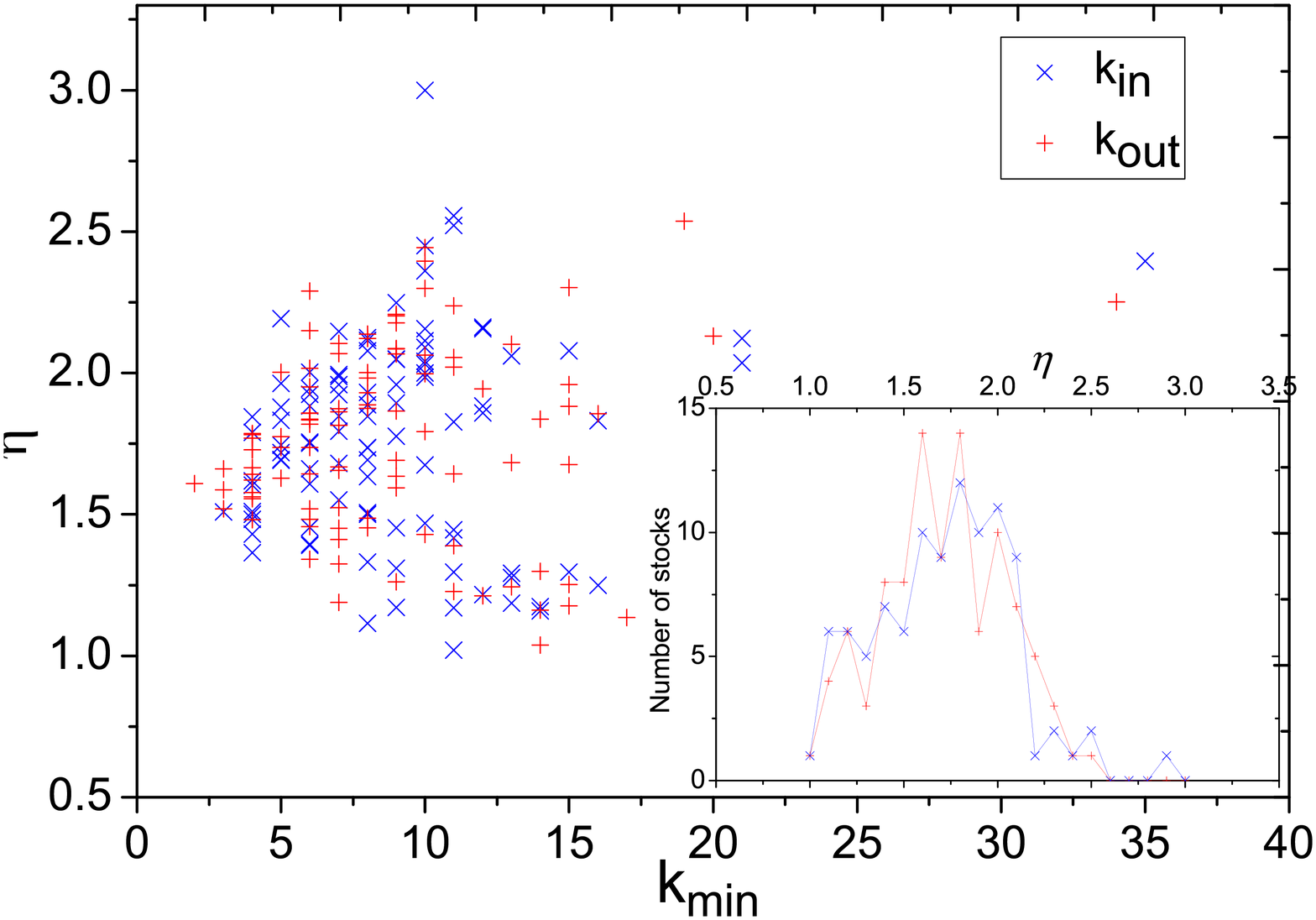}
    \end{minipage}}%
\caption{(a)Cumulative probability distribution of out-degree and in-degree on a log-log scale for 5 randomly chosen stocks. The solid lines are the best fits to the power-law tails. (b)Scatter plot of the pairs($k_{min}$, $\eta$) for 100 stocks. Each point represents the lower boundary of power-law, $k_{min}$, and the exponent. The inset shows the empirical frequency of $\eta$. The mean values of the exponents are $\eta_{out}$=$1.7658\pm0.33$,$\eta_{in}$=$1.8045\pm0.37$.\label{degree}}%
\end{figure}

Figure \ref{degree}(a) illustrates the cumulative probability distribution of out-degrees and in-degrees for five randomly chosen stocks. We can see that there is a power-law decay when \emph{k} is larger than $k_{min}$, such that:
\begin{equation}
    P_{out}(k)\sim{k^{-\eta_{out}}},\texttt{ if $k \geq k_{min}$.}
\end{equation}
\begin{equation}
    P_{in}(k)\sim{k^{-\eta_{in}}}, \texttt{if $k \geq k_{min}$.}
\end{equation}

In order to confirm our observations of these power-law distributions, we adopt the least square estimate to fit the curve and test the goodness of fit with the Kolmogorov-Smirnov (KS) statistic. The calibrating parameters $k_{min}$ and $\eta$ are substituted into the KS test procedure to test the goodness of fit. The null hypothesis $H_{0}$ for our KS test is that the data ($k$ $\geq k_{min}$) are drawn from a power-law distribution. In~\cite{Newman}, a power-law fitting approach for estimating the power-law exponent of empirical data and the lower bound of the power-law behavior is proposed, which combines maximum-likelihood fitting methods with goodness-of-fit tests based on the KS statistic and likelihood ratios. We calibrate the empirical data by this approach and find that the null hypothesis $H_{0}$ cannot be rejected at the significant level of 0.01 for all the 100 non-manipulated stocks. Figure \ref{degree}(b) depicts the bivariate distribution of pairs ($k_{min}$, $\eta_{out}$) and ($k_{min}$, $\eta_{in}$). Each node represents the lower boundary of power-law tail $k_{min}$ and the estimated power-law exponents $\eta_{out}$ and $\eta_{in}$. As shown in the inset of figure \ref{degree}(b) the average values are $\eta_{out}$=$1.7658\pm0.33$ and $\eta_{in}$=$1.8045\pm0.37$. There are 75\% stocks whose $\eta_{out}$ values are well within the L$\acute{e}$vy-stable regime and 73\% stocks whose $\eta_{in}$ values are located in the L$\acute{e}$vy-stable regime. We also notice that the degree distribution of the directed trading network also falls in the L$\acute{e}$vy-stable regime~\cite{Zhouweixing10}. The degree distributions show that a few investors have many connections and most investors only have a few links.

\subsection{The node strength distribution}

In order to understand the investors' trading behavior for one stock, we analyze the node strength of the trading network. The node strength is defined as the sum of all edge weights of the node. We define the weight on an edge $e_{ij}$ as the total volume exchanged from \emph{i} to \emph{j}. We investigate three types of strength: the in-strength $S_{in}(i)=\sum_{j} w_{ji}$, the out-strength $S_{out}(i)=\sum_{j} w_{ij}$, and the total strength $S(i)=S_{in}(i)+S_{out}(i)$, where $w_{ij}$ is the weight on link i$\rightarrow$j. The in-strength of a node is the total volume bought by the corresponding trader in a certain period , the out-strength means the total volume that an investor sold and the total strength is equal to the total trading volume exchanged by an investor in a certain period. We analyze the trading volume of each investor for one stock in a year to find the statistical properties of the investors' trading behavior.

After calculating the cumulative distribution of 100 stocks, we find that the strength distributions exhibit power-law decay in the tails when $s>s_{min}$:
\begin{equation}
   P_{in}(s)\sim{s^{-\gamma_{in}}},
   P_{out}(s)\sim{s^{-\gamma_{out}}},
   P(s)\sim{s^{-\gamma}}.
\end{equation}

\begin{figure}
\centering
\subfigure[]{
    \label{fig:mini:subfig:a}
    \begin{minipage}[b]{0.5\textwidth}
    \centering
    \includegraphics[width=\textwidth]{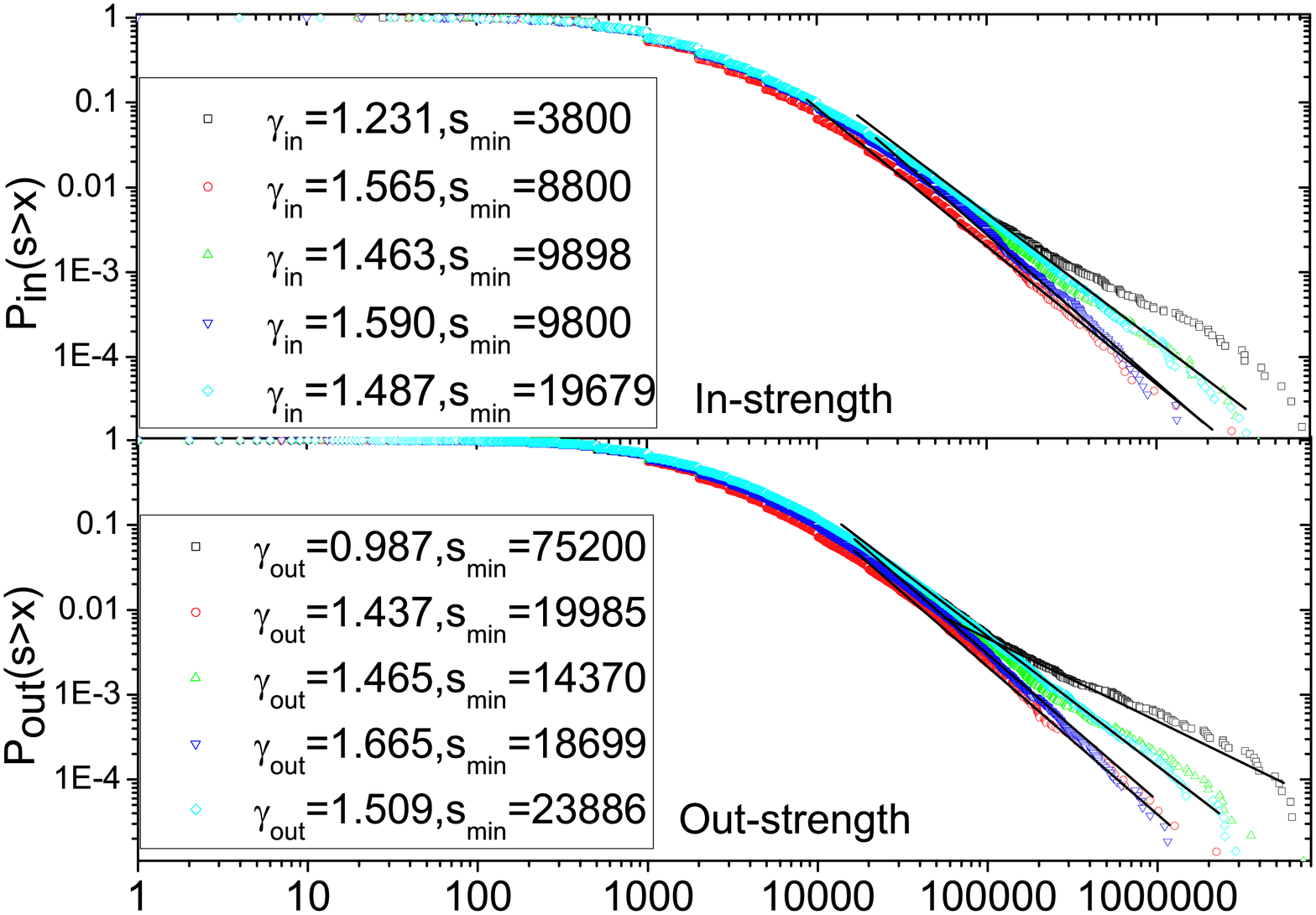}
    \end{minipage}}%
\subfigure[]{
    \label{fig:mini:subfig:b}
    \begin{minipage}[b]{0.5\textwidth}
    \centering
    \includegraphics[width=\textwidth]{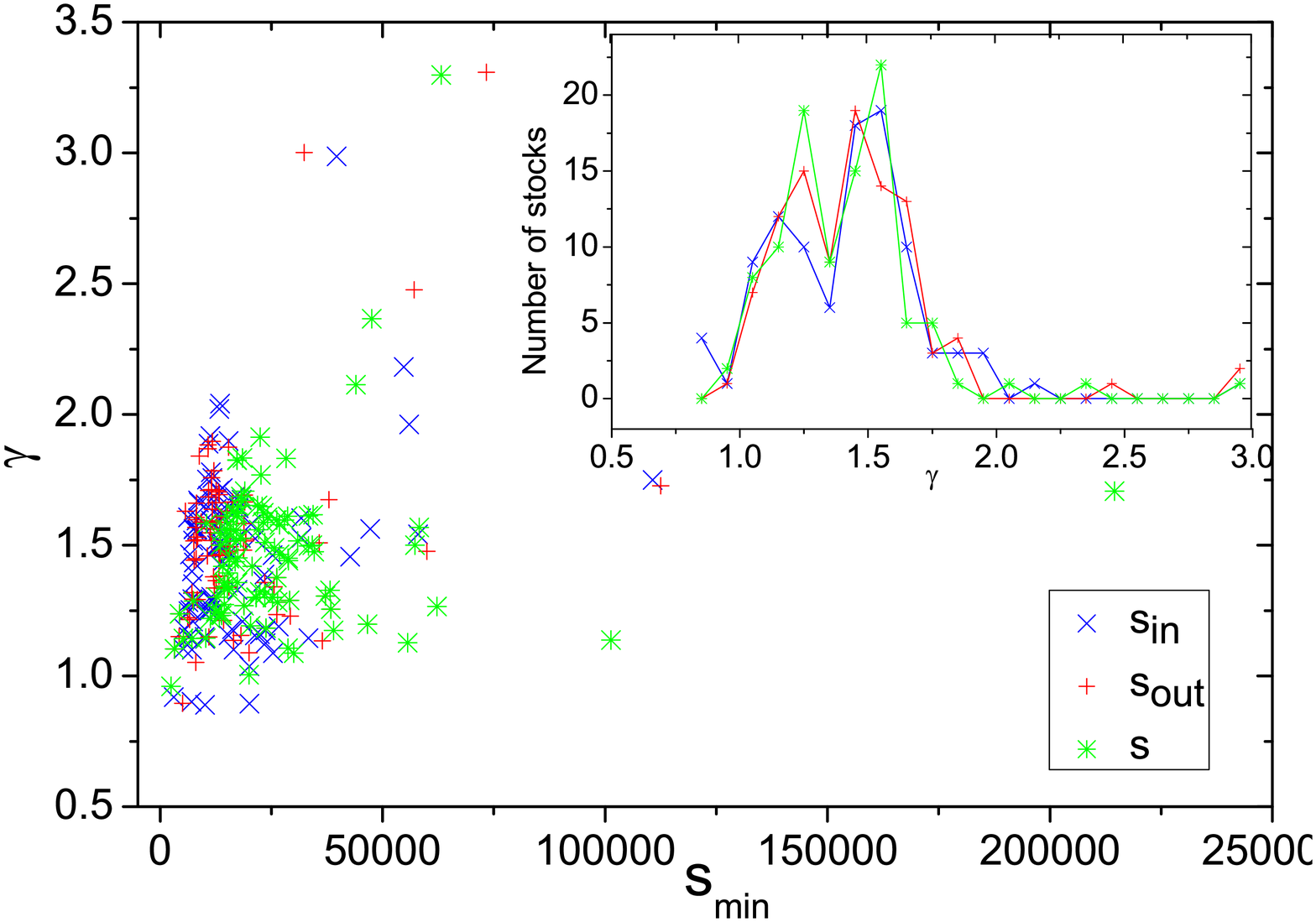}
    \end{minipage}}%
\caption{(a)Cumulative probability distributions of the out-strength and in-strength on a log-log scale for five randomly chosen stocks. The solid lines are the best fits to the power-law tails. (b)Scatter plot of the pairs($k_{min}$, $\gamma$) for 100 stocks. Each point represents the lower boundary of the power-law $s_{min}$ and the exponent. The inset shows the empirical frequency of $\gamma$. The mean values of the exponents are $\gamma_{out}$=$1.4987\pm0.33$,$\gamma_{in}$=$1.4660\pm0.31$ and $\gamma$=$1.4672\pm0.30$.\label{strength}}%
\end{figure}

We use the same strategy as was utilized in Section 3.2 to confirm our observations of power-law distributions. The KS test is implemented on the calibrating results to check the goodness of fit. Our null hypothesis $H_{0}$ is that the strength can be well modeled by a power-law distribution. At a significance level of 0.01, we find that the null hypothesis $H_{0}$ cannot be rejected for all 100 stocks for in-strength $S_{in}$, out-strength $S_{out}$ and total strength S. Figure \ref{strength}(a) illustrates the in-strength and out-strength cumulative distributions for five randomly chosen stocks, and the solid lines are the best fits to the power-law distributions. Figure \ref{strength}(b) depicts the bivariate distribution of pairs($s_{min}$,$\gamma$). Each node is associated with the lower bound of the power-law tail $s_{min}$ and the estimated power-law exponents $\gamma$. The inset shows the frequency of exponent $\gamma$, and the average values are $\gamma_{out}$=$1.4987\pm0.33$, $\gamma_{in}$=$1.4660\pm0.31$ and $\gamma$=$1.4672\pm0.30$. From the Figure\ref{strength}(b), we find that power-law exponents $\gamma$ lie in [0,2] for the in-strength distribution, out-strength distribution and total strength distribution.

\section{Trading activity of manipulated stocks}

The first researchers who studied manipulation were Allen, Gale and Jarrow. They studied the history of the stock price manipulation and classified the manipulations into three types: information-based manipulation, action-based manipulation and trade-based manipulation. Today, the most common types of manipulation are information-based manipulation and trade-based manipulation. In information-based manipulation, manipulators spread rumors and false information to influence the stock prices. In trade-based manipulation, manipulators engage in fraudulent trading to create an image of an active market~\cite{Allen92,Jarrow92}. The manipulated stocks in our data set are published by China Securities Regulatory Commission(CSRC) for trade-based manipulation. In this paper, we investigate the trading network for the analysis of trade-based manipulation only.

\subsection{The degree and strength distribution of manipulated stocks}

After achieving an understanding of the statistical properties of stock trading networks, we want to know whether the manipulated stocks exhibit the same phenomenon as non-manipulated stocks. If the manipulated stocks are different from non-manipulated ones, how different are they? In order to answer these questions, we analyze eight manipulated stocks published by CSRC. In our data set, five of these eight stocks have the manipulated period through a whole year and the remaining three stocks' manipulated periods are from Jan 2004 to Sep 2004.

In order to compare the differences between non-manipulated stocks and manipulated stocks under the same conditions, we take the following method to select a set of non-manipulated stocks as reference stocks. First, for each manipulated stock, the reference stocks are the non-manipulated stocks which have the same capitalization and belong to the same industry sector. Second, we compute the degree distribution and strength distribution of the trading networks for each manipulated stock during the manipulation period and we also compute the same statistics of the reference stocks during the same periods. Third, for each stock we fit and test the power-law hypothesis using the methods described in the previous sections, and for each manipulated stock we take the average value of its reference stocks as the reference value.

\begin{figure}
\centering
\subfigure[]{
    \label{fig:mini:subfig:a}
    \begin{minipage}[b]{0.5\textwidth}
    \centering
    \includegraphics[width=\textwidth]{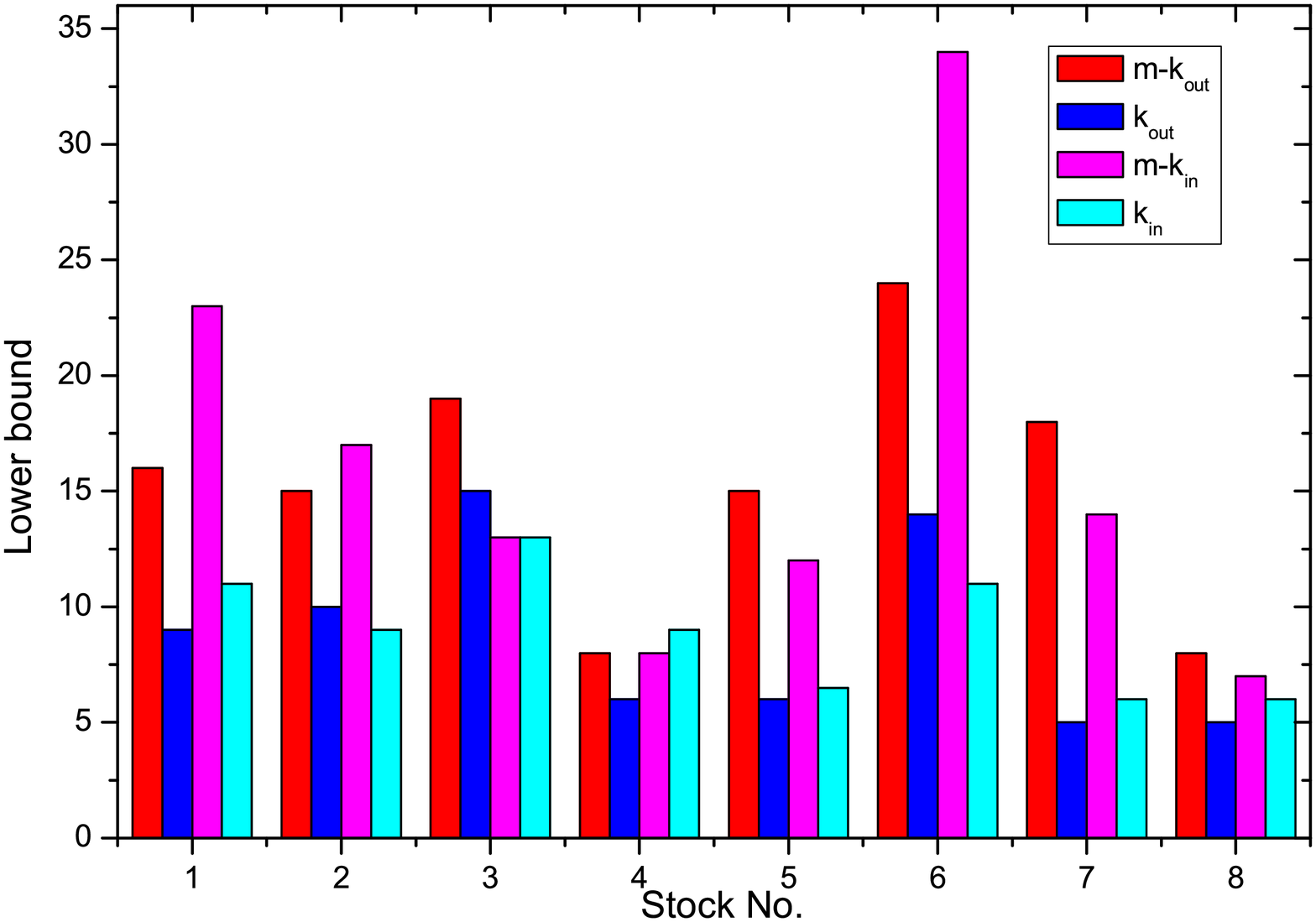}
    \end{minipage}}%
   \vspace{0.1in}
\subfigure[]{
    \label{fig:mini:subfig:b}
    \begin{minipage}[b]{0.5\textwidth}
    \centering
    \includegraphics[width=\textwidth]{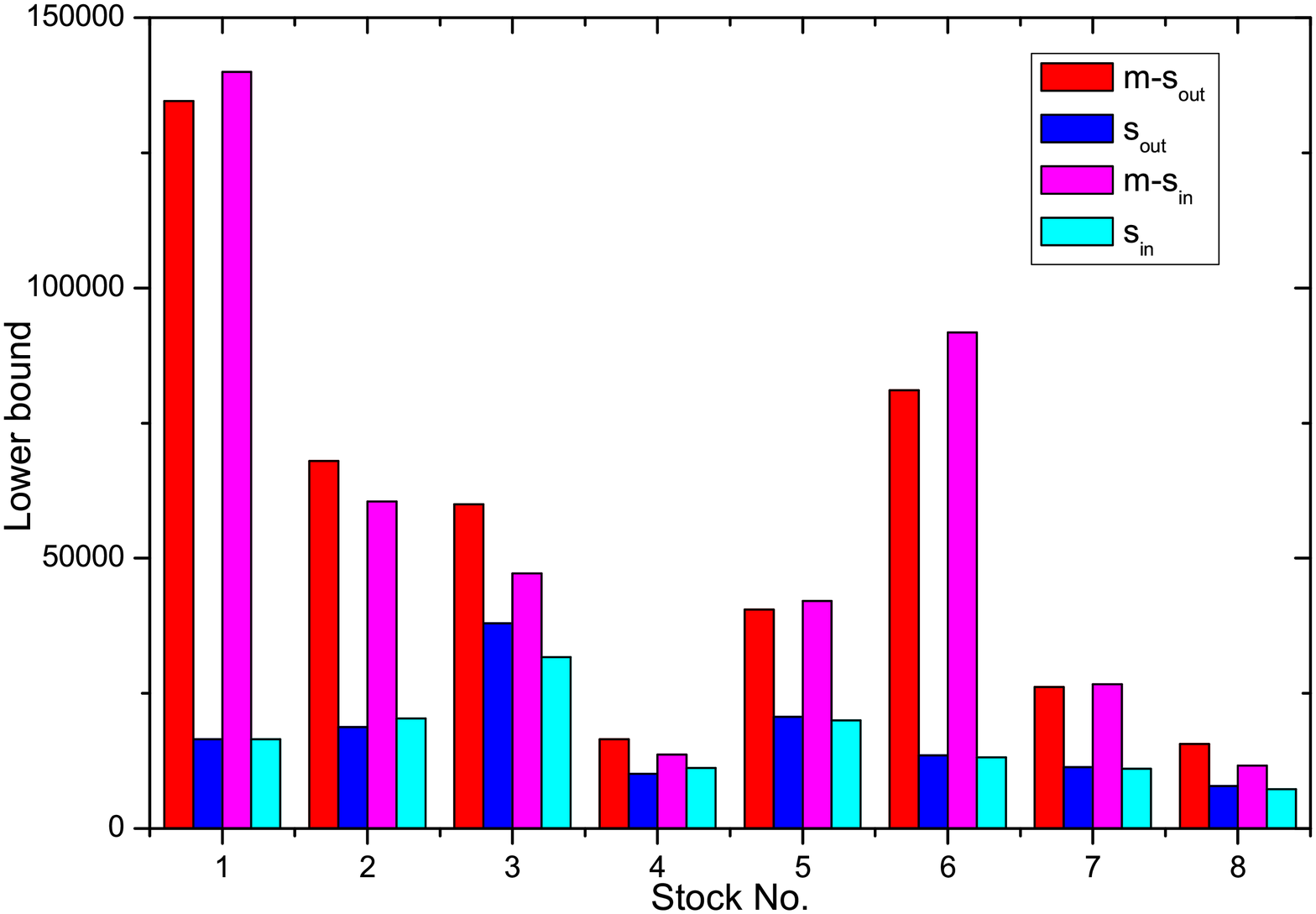}
    \end{minipage}}%
    \vspace{0.1in}
\subfigure[]{
    \label{fig:mini:subfig:c}
    \begin{minipage}[b]{0.5\textwidth}
    \centering
    \includegraphics[width=\textwidth]{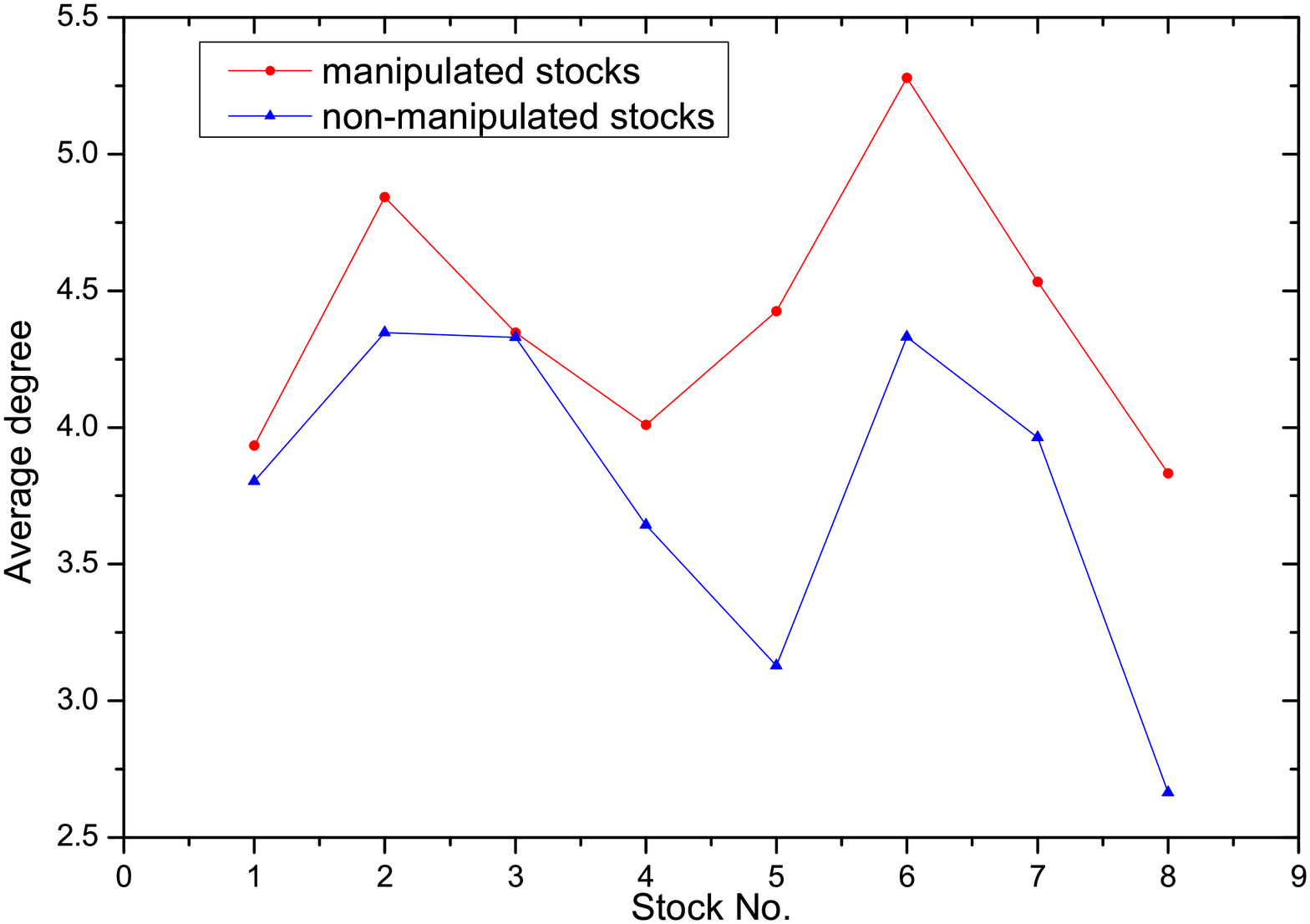}
    \end{minipage}}%
\caption{(a)Histogram of the lower bounds of the power-law fit of the degree distribution for eight manipulated stocks and non-manipulated stocks. (b)Histogram of the lower bounds of the power-law fit of the strength distribution for eight manipulated stocks and non-manipulated stocks. (c)The average degrees of trading networks. The red bars represent the statistics of eight manipulated stocks and the blue ones represent the reference values . \label{lowerbound}}%
\end{figure}

In Figure \ref{lowerbound}(a) and (b), for all power-law distributions mentioned above, the lower bounds of each power-law fit for manipulated stocks are much larger than the reference values(except in-degree lower bound of the fourth stock). From Figure \ref{lowerbound}(c), it can be seen that the average degrees of the trading networks of manipulated stocks are larger than those for non-manipulated stocks. This means that there are more transactions between traders of the manipulated stocks. This phenomenon is attributed to some malpractices involving a group of traders acting and trading together to achieve a specific effect on the volume of a target security. In fact, manipulators trade among themselves in order to artificially increase the price and volume of a stock for the purpose of attracting other investors to buy the stock and they have a heavy trading volume among themselves. Manipulators earn a profit and investors incur losses. The anomaly in these distributions means that the manipulators are disturbing the market order.

\subsection{Price return and trader number}

\begin{figure}
\centering
\subfigure[]{
    \label{fig:mini:subfig:a}
    \begin{minipage}[b]{0.5\textwidth}
    \centering
    \includegraphics[width=\textwidth]{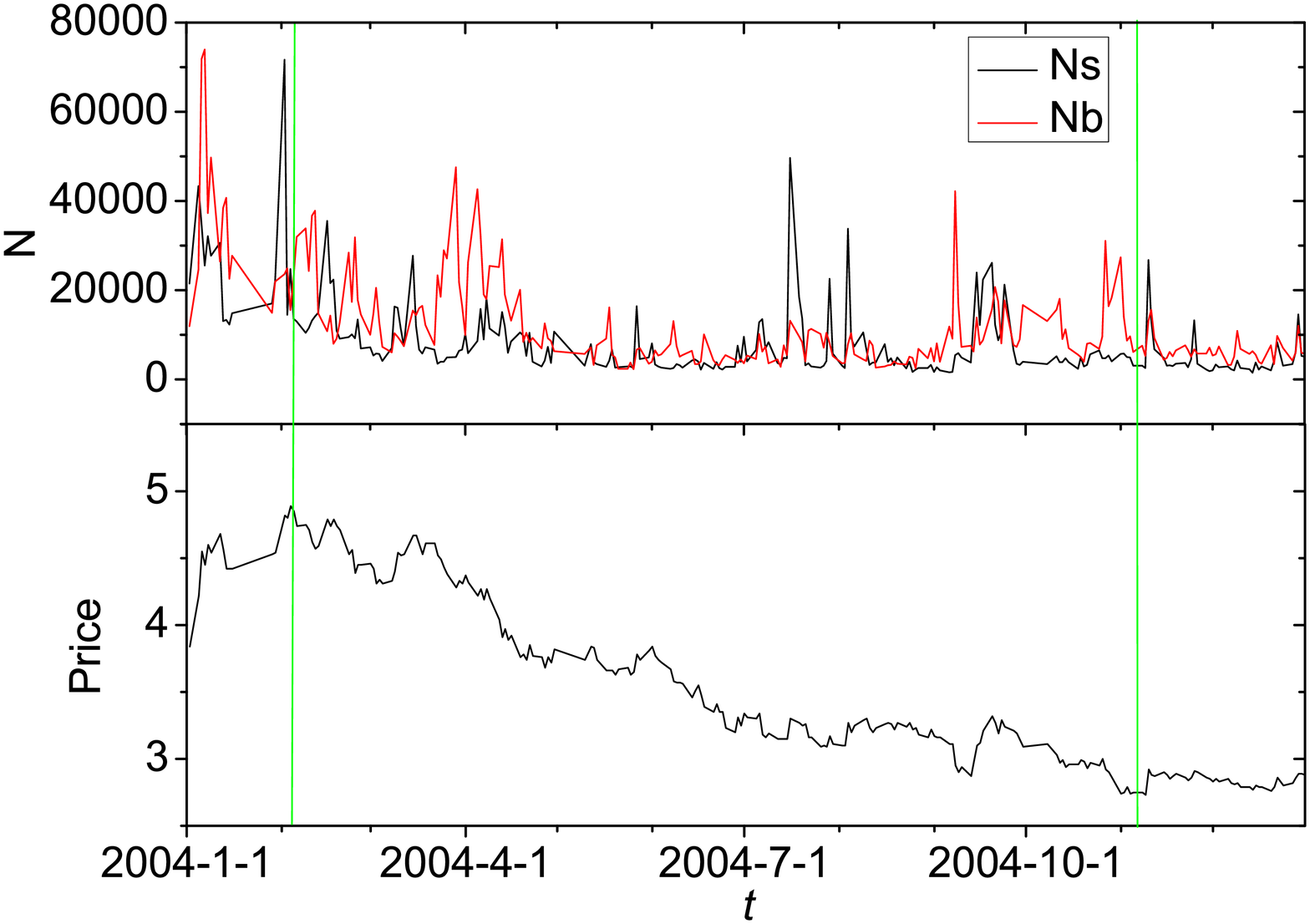}
    \end{minipage}}%
\subfigure[]{
    \label{fig:mini:subfig:b}
    \begin{minipage}[b]{0.5\textwidth}
    \centering
    \includegraphics[width=\textwidth]{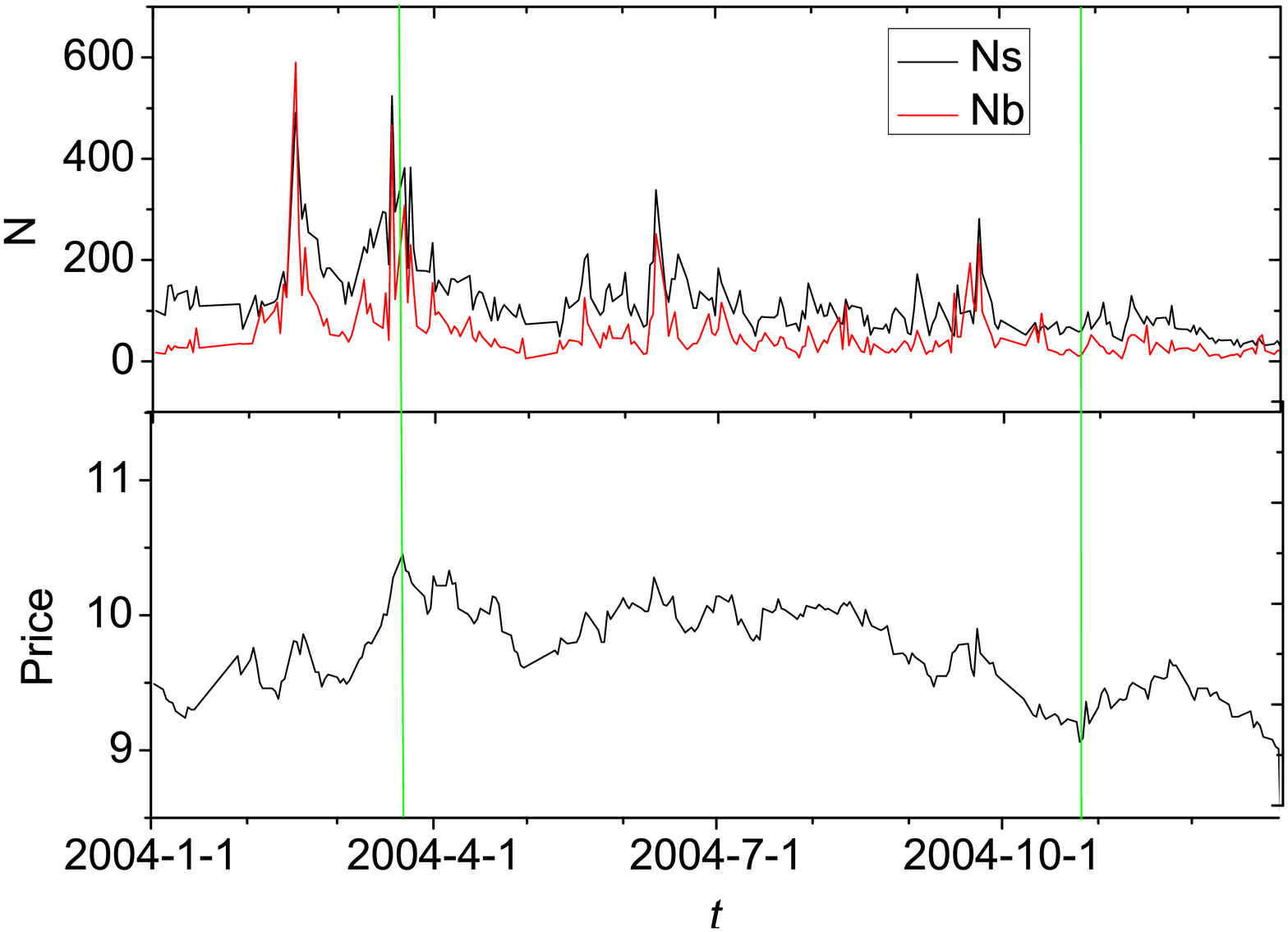}
    \end{minipage}}%
\caption{Evolution of $N_{s}$ and $N_{b}$ with the trading days. The vertical lines separate the highest price and the lowest price. The price is presented in the low panel. (a) depicts a randomly chosen non-manipulated stock and (b) shows a manipulated stock.\label{price}}%
\end{figure}

After investigating the statistical properties of a manipulated trading network, we study how price change correlates with the trading activities. First, we consider how the number of traders changes with price. We plot the number $N_{b}$ of traders who buy shares and the number $N_{s}$ of traders who sell shares with respect to the trading day \emph{t} in the top panel of Figure \ref{price} and the stock prices are also plotted in the bottom panel.

Figure \ref{price} illustrates the evolution of the daily prices, $N_{s}$ and $N_{b}$, from which one can see that $N_{s}$ and $N_{b}$ change synchronously with the stock prices for both non-manipulated stock and manipulated stock. For non-manipulated stock (Figure \ref{price}(a)), when the price goes up, $N_{s}>N_{b}$; otherwise $N_{b}>N_{s}$. Compared with the non-manipulated stocks, the manipulated stock (Figure \ref{price}(b)) behaves differently. For this stock, $N_{s}$ is always greater than $N_{b}$ no matter whether price rises or falls.

As we know, prices of stocks must be determined by the listed companies' condition without any interference, and the price of a certain stock can affect the investors' decision to buy or to sell. In order to obtain more profit, investors buy the stocks at low prices and sell them at high prices, which is called "buy low and sell high". When a buyer submits a big bid order, there will be a great number of retail investors who sell their shares. There are more buyer-initiated trades than seller-initiated trades and the price will go up. However, manipulators employ different methods to influence the price of the targeted stocks, and the number of traders may not reflect the price change.

\begin{figure}[!htb]
\includegraphics[width=\textwidth]{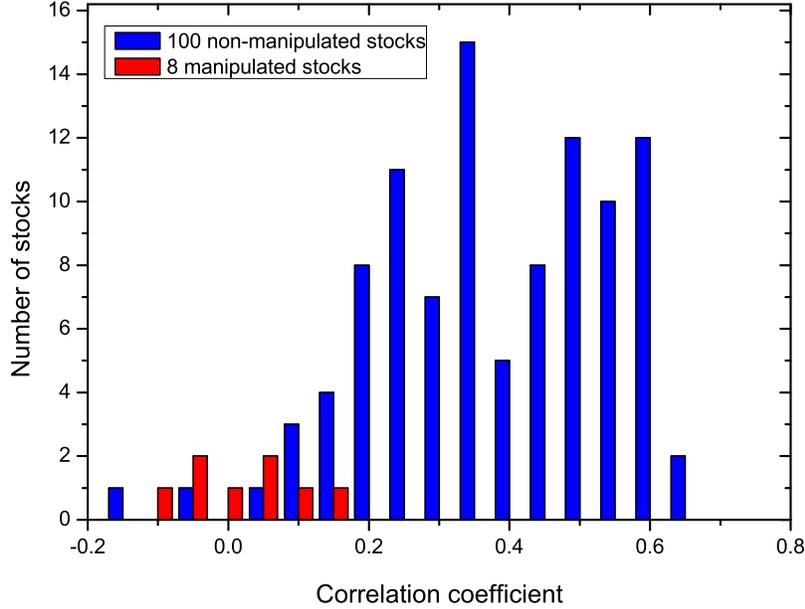}
\caption{Histogram of the coefficient of correlation between \emph{r} and the price return. The blue bars represent 100 non-manipulated stocks and the red bars is the eight manipulated stocks\label{correlation}}
\end{figure}

Then we study the correlation between price return and the number of traders for each non-manipulated stock and manipulated stock. Each return series was evaluated through the logarithmic change of the corresponding price series, \emph{pr(t)=$\ln$P(t)-$\ln$P(t-1)}~\cite{Stanley99}, in which \emph{P(t)} denotes the average price on day $\emph{t}$. The seller-buyer ratio \emph{r} is defined as the ratio of $N_{s}$ to $N_{b}$. Next, we calculate the correlation coefficient of the price return and \emph{r}. This is illustrated in Figure \ref{correlation}. From Figure \ref{correlation}, we find that \emph{r} is positively correlated to \emph{pr(t)} and the correlation coefficients are greater than 0.2 for most of the non-manipulated stocks; however, for the manipulated stocks they are all less than 0.2.

In trade-based manipulation, manipulators engage in fraudulent trading to create an image of an active market and attract the other investors to buy the stock~\cite{Allen92}. First, a manipulator controls hundreds of accounts in order to purchase or sell a large amount of shares at the same time in order to influence the stock price. Second, money transfers from one or several satellite accounts that are also controlled by the fraudulent trader to a central account. A manipulator trade shares from one account to another without changing ownership and this fraudulent trading leads to an image of active market. Therefore, price fluctuations of manipulated stocks do not reflect the demand fluctuations.

\section{Conclusion}
In this paper, we investigate the statistical properties of investor's trading behavior by means of network analysis based on the transaction data for 108 stocks during the whole year 2004.

For each stock, we construct a trading network in which the nodes represent investors and each transaction is translated into a directed edge drawn from a seller to a buyer with the total trading volume as its weight. We find that the degree distributions follow a power law and the node strength distributions display power laws in the tails.

Furthermore, we examine the statistics of the manipulated stocks. We choose the non-manipulated stocks which have the same capitalization and industry sector with the manipulated stocks. By comparing the degree and strength distributions between non-manipulated stocks and manipulated stocks, we find that manipulated stocks have a higher lower bound of the power-law tail, and higher average degree of the trading network than non-manipulated ones. By analyzing the correlation between the price return and seller-buyer ratio, we find that manipulated stocks have a lower correlation than non-manipulated stocks.

A trading network provides a way to trace shares and cash flows in financial markets. However, the transaction records are generated by the trading system automatically according to the principle of price-time priority rather than from the direct results of the negotiation among users and spontaneous behavior. How to combat such a drawback is an issue for future work.

\section*{Acknowledgments}
This work was funded by the National Natural Science Foundation of China under grant numbers 60873245, 60933005, and 60803123. This work was also partly founded by the National High-Tech R\&D (863) Program of China under grant number 2010AA012500.


\bibliographystyle{model1a-num-names}



\end{document}